\documentclass[final,5p,times,twocolumn,authoryear]{elsarticle}

\usepackage{amssymb}

\usepackage{savesym}
\savesymbol{tablenum}
\usepackage{siunitx}
\restoresymbol{SIX}{tablenum}

\usepackage{tipa}
\usepackage{amsmath}
\usepackage{ulem}

\usepackage{mathcomp}

\DeclareMathOperator\erf{erf}

\usepackage{lineno}

\journal{Icarus}

\begin{document}

\begin{frontmatter}



\title{Oxygen Isotope Exchange Between Molten Silicate Spherules and Ambient Water Vapor with Nonzero Relative Velocity: Implication for Chondrule Formation Environment}


\author[inst1]{Sota Arakawa}
\author[inst2]{Daiki Yamamoto}
\author[inst3]{Takayuki Ushikubo}
\author[inst4]{Hiroaki Kaneko}
\author[inst5]{Hidekazu Tanaka}
\author[inst1]{Shigenobu Hirose}
\author[inst4]{Taishi Nakamoto}

\affiliation[inst1]{organization={Yokohama Institute for Earth Sciences, Japan Agency for Marine-Earth Science and Technology},
            addressline={3173-25 Showa-machi, Kanazawa-ku}, 
            city={Yokohama},
            postcode={236-0001}, 
            country={Japan}}

\affiliation[inst2]{organization={Department of Earth and Planetary Sciences, Kyushu University},
            addressline={744, Motooka, Nishi-ku}, 
            city={Fukuoka},
            postcode={819-0395}, 
            country={Japan}}

\affiliation[inst3]{organization={Kochi Institute for Core Sample Research, Japan Agency for Marine-Earth Science and Technology},
            addressline={200 Monobe-otsu}, 
            city={Nankoku, Kochi},
            postcode={783-8502}, 
            country={Japan}}

\affiliation[inst4]{organization={Department of Earth and Planetary Sciences, Tokyo Institute of Technology},
            addressline={2-12-1 Ookayama, Meguro}, 
            city={Tokyo},
            postcode={152-8550}, 
            country={Japan}}

\affiliation[inst5]{organization={Astronomical Institute, Graduate School of Science, Tohoku University},
            addressline={6-3 Aramaki, Aoba-ku}, 
            city={Sendai},
            postcode={980-8578}, 
            country={Japan}}

\begin{abstract}
Oxygen isotope compositions of chondrules reflect the environment of chondrule formation and its spatial and temporal variations.
Here, we present a theoretical model of oxygen isotope exchange reaction between molten silicate spherules and ambient water vapor with finite relative velocity.
We found a new phenomenon, that is, mass-dependent fractionation caused by isotope exchange with ambient vapor moving with nonzero relative velocity.
We also discussed the plausible condition for chondrule formation from the point of view of oxygen isotope compositions.
Our findings indicate that the relative velocity between chondrules and ambient vapor would be lower than several $100~\si{m}~\si{s}^{-1}$ when chondrules crystallized.
\end{abstract}




\end{frontmatter}



\section{Introduction} \label{sec:intro}

Chondrules are igneous ferromagnesian silicate spherules and they are thought to be formed via transient and high-temperature processes in the gaseous solar nebula \citep[e.g.,][]{2018crpd.book...57J, 2018crpd.book...11K}.
They are contained within chondrites, which are the most common type of meteorites, as a major component \citep[15--80\% in volume except for CI chondrites;][]{2007AREPS..35..577S}.
Therefore, their mineralogical and chemical properties are the key to understanding the physicochemical condition of planet(esimal)-forming environments.

Oxygen isotope compositions of chondrules have evolved via isotope exchange reaction with ambient vapor, and those would reflect the composition of the chondrule-forming environment (e.g., the dust-to-gas and water-to-rock ratios) and kinetic effects \citep[e.g.,][]{2018crpd.book..192T, 2021GeCoA.313..295P}.
\citet{2012GeCoA..90..242U} measured the oxygen isotope compositions of chondrules in Acfer 094, one of the least thermally or aqueously altered carbonaceous chondrites whose petrologic type is 3.00 \citep{2008M&PS...43.1161K}, and they reported the following notable features: (1) phenocrysts and mesostasis in the same chondrule have similar oxygen isotope compositions, (2) porphyritic chondrules frequently host relict olivine grains that are distinguished by the mass-independent difference in oxygen isotope compositions \citep[see also][]{2004GeCoA..68.3599K, 2005GeCoA..69.3831K}, and (3) relict olivine grains in the same chondrule are either all $^{16}{\rm O}$-enriched or all $^{16}{\rm O}$-depleted compared to chondrule phenocrysts, i.e., $^{16}{\rm O}$-enriched and $^{16}{\rm O}$-depleted relict olivine grains are exclusively present in one chondrule.\footnote{
Although this is mostly true, some exceptions were reported by \citet{2019PNAS..11623461M}.
}
These features are commonly observed in various chondrites \citep[e.g.,][]{2018crpd.book..192T, 2018E&PSL.496..132M, 2020GeCoA.282..133S, LIBOUREL2023102} and regarded as the evidence of the efficient oxygen isotopic exchange between (molten) chondrules and the ambient vapor \citep[e.g.,][]{2010GeCoA..74.6610K}.
Evaporation and recondensation of silicate would also cause the variation of the oxygen isotope composition of chondrules \citep[e.g.,][]{2004GeCoA..68.3943A, 2012M&PS...47.1209N}.

In carbonaceous chondrites, the oxygen isotope compositions of chondrules are distributed along a line with slope $\sim 1$, which is called the primitive chondrule minerals (PCM) line, in the three-oxygen isotope diagram \citep{2012GeCoA..90..242U, 2022ChGeo.608l1016Z}.
This trend cannot be explained by the mass-dependent isotopic fractionation associated with physical and chemical processes that makes a variation along a line with slope $\sim 0.52$ \citep[e.g.,][]{1983E&PSL..62....1C}, and it is usually interpreted as the results of mixing between isotopically distinct reservoirs \citep[e.g.,][]{2018crpd.book..192T}.
Therefore, oxygen isotope exchange reaction between chondrule precursors and ambient vapor have a potential to explain the variation of oxygen isotope compositions of chondrules.

One of the possible reservoirs of $^{16}{\rm O}$-depleted materials is ${\rm H}_{2}{\rm O}$ ice in the outer solar system as products of self-shielding of carbon monoxide in the protosolar molecular cloud \citep[e.g.,][]{2004Sci...305.1763Y} or the solar nebula \citep[e.g.,][]{2005Natur.435..317L}.
\citet{2021GeCoA.314..108Y} performed laboratory experiments to determine the oxygen isotope exchange kinetics between calcium--aluminum-rich inclusion (CAI) analogue melt and water vapor.
They found that the oxygen isotope exchange efficiency\footnote{The isotope exchange efficiency is mathematically identical to the condensation coefficient \citep[e.g.,][]{2011ApJ...736...16T, 2015ApJS..218....2T}.} on the melt surface is $\beta \sim 0.3$ in colliding water molecules ($\beta$ is defined later in Equation (\ref{eq:beta}); see Section \ref{sec:influx} and Figure \ref{fig3}).
Using the $\beta$ value obtained from laboratory experiments, we can calculate the temporal evolution of oxygen isotope ratios from the Hertz--Knudsen equation \citep[e.g.,][]{2018ApJ...865...98Y, 2020M&PS...55.1281Y, 2021GeCoA.314..108Y}.
It should be noted that the Hertz--Knudsen equation is only applicable when the relative velocity between silicate melt and the bulk motion of water vapor is zero.
However, this condition might be broken in chondrule formation environments if silicate melts formed by dynamic processes such as shock waves.

The formation mechanisms of chondrules are still under debate.
One of the leading candidates is shock-wave heating \citep[e.g.,][]{2001Icar..153..430I, 2018crpd.book..375M} in the gaseous solar nebula.
Other well-researched scenarios are radiative heating by lightning \citep[e.g.,][]{2018A&A...609A..31J, 2023ApJ...947...15K} and planetesimal collisions \citep[e.g.,][]{2018crpd.book..343J, 2018crpd.book..361S}.
The shock-wave heating model could cause a large relative velocity between chondrules and ambient vapor.
Chondrules formed via planetesimal collisions might also have a large relative velocity.
The presence of compound chondrules is also regarded as evidence of nonzero velocities of chondrules \citep[e.g.,][]{1996AMR.....9..208S, 2016Icar..276..102A, 2021GeCoA.296...18J}.

In this study, we develop a theoretical model of isotope exchange reaction between molten spherules and ambient vapor with nonzero relative velocities.
We derived a modified equation of the flux of ambient vapor that is applicable for moving spheres (Section \ref{sec:influx}), and we revisited the relation between the influx and efflux in isotope exchange reactions based on the concept of the isotope fractionation at the equilibrium (Section \ref{sec:efflux}).
We found that huge mass-dependent fractionation would be caused by isotope exchange with ambient vapor when the relative velocity is comparable or larger than the most probable speed of reacting molecules (Section \ref{sec:results}).
Our findings indicate that the relative velocity between chondrules and ambient vapor would be lower than several $100~\si{m}~\si{s}^{-1}$ when chondrules crystallized.
Discussions on the chondrule-forming environments are present in Section \ref{sec:discussion}.

\section{Model} \label{sec:model}

\begin{figure}
\begin{center}
\includegraphics[width=\columnwidth]{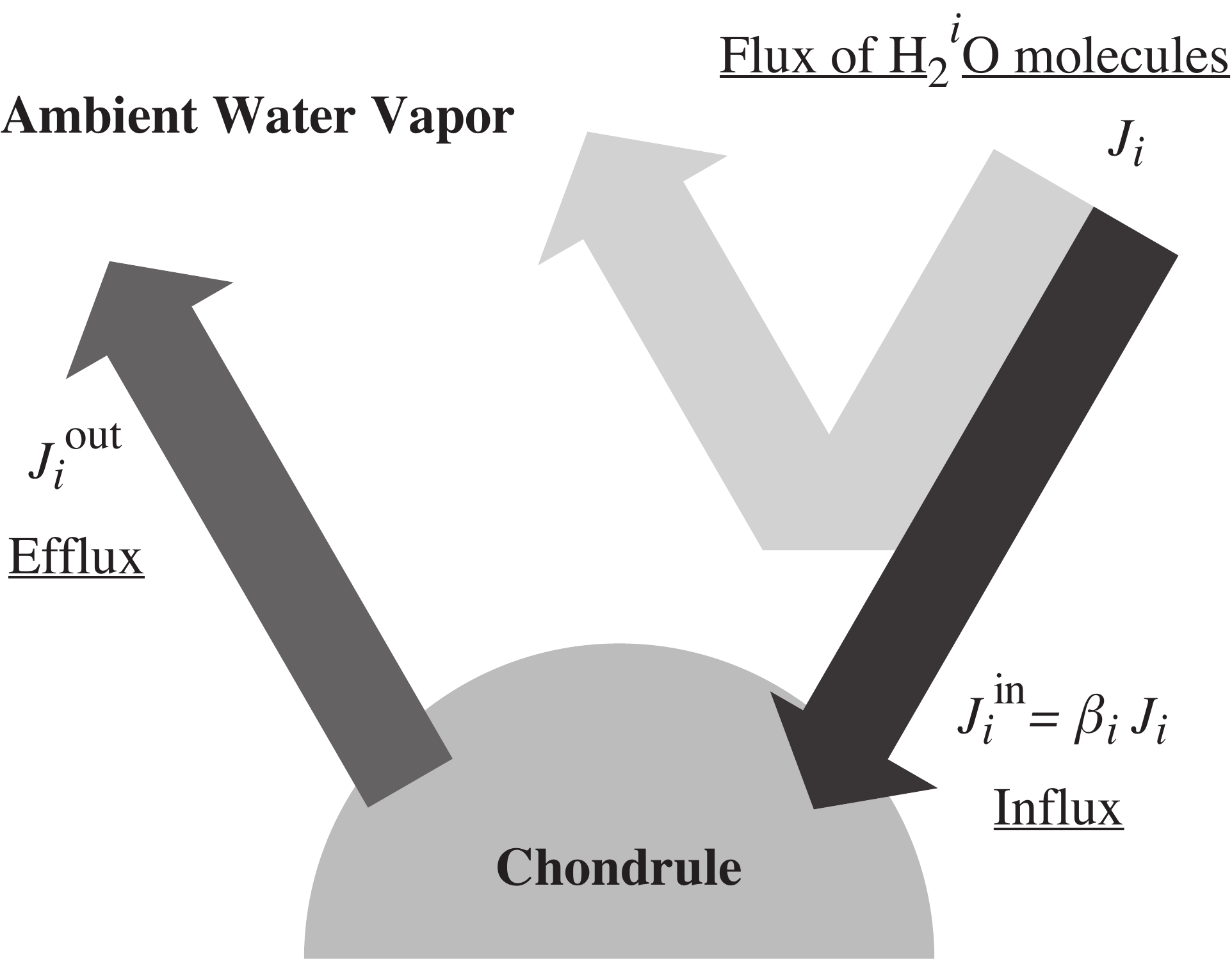}
\end{center}
\caption{
Schematic of the influx and efflux of water molecules at the molten chondrule surface.
The influx of ${\rm H}_{2} {}^{i}{\rm O}$ molecules which can react with a chondrule, $J_{i}^{\rm in}$, is given by $J_{i}^{\rm in} = \beta_{i} J_{i}$, where $J_{i}$ is the flux of ${\rm H}_{2} {}^{i}{\rm O}$ molecules and $\beta_{i}$ is the isotope exchange efficiency.
We note that $J_{i}$ depends on relative velocity of the chondrule and the ambient vapor, $v_{\rm rel}$ (see Equation (\ref{eq:J})).
The efflux of ${\rm H}_{2} {}^{i}{\rm O}$ molecules after isotope exchange reaction is $J_{i}^{\rm out}$.
We assumed that $J_{i}^{\rm out} / J_{16}^{\rm out}$ is a function of the oxygen isotope composition of the chondrule (see Equation (\ref{eq:efflux_i/16})).
}
\label{fig3}
\end{figure}

The oxygen isotope exchange reaction is controlled by the balance between the influx and efflux of ambient vapor.
As a first step, we consider the simplest situation, i.e., molten chondrules react only with water vapor.
Figure \ref{fig3} is the schematic of the influx and efflux of water molecules at the molten chondrule surface.
In this section, we describe the theoretical model of the influx and efflux which can be applied to chondrules moving with finite velocity.

It should be noted that evaporation of silicate melt would also cause isotope fractionation.
We will discuss the impacts of evaporation and recondensation on the oxygen isotope ratio of silicate melt in Section \ref{sec:evaporation}.

\subsection{Influx of water molecules} \label{sec:influx}

In this section, we derive the influx of water molecules under the situation that the chondrule moves in gas with relative velocity $v_{\rm rel}$.
When $v_{\rm rel} = 0$, the flux of colliding water molecules (per unit area) at the chondrule surface is given by the Hertz--Knudsen equation:
\begin{equation}
J_{0} = \frac{P_{\rm water}}{\sqrt{2 \pi m k_{\rm B} T}},
\end{equation}
where $P_{\rm water}$ is the partial pressure of ambient water vapor, $m$ is the molecular weight of water, $k_{\rm B}$ is the Boltzmann constant, and $T$ is the temperature of ambient water vapor.
The distribution of the normal component of the velocity of water molecules, $v_{\rm n}$, is given by the Maxwell distribution:
\begin{equation}
{f ( v_{\rm n} )} = \sqrt{\frac{m}{2 \pi k_{\rm B} T}} \exp{\left( - \frac{m {v_{\rm n}}^{2}}{2 k_{\rm B} T} \right)},
\end{equation}
and, by definition,
\begin{equation}
\int_{- \infty}^{+ \infty} {f ( v_{\rm n} )} ~{\rm d}v_{\rm n} = 1.
\end{equation}
Here, $v_{\rm n}$ is counted positively when water molecules approach to the chondrule surface.

\begin{figure}
\begin{center}
\includegraphics[width=0.5\columnwidth]{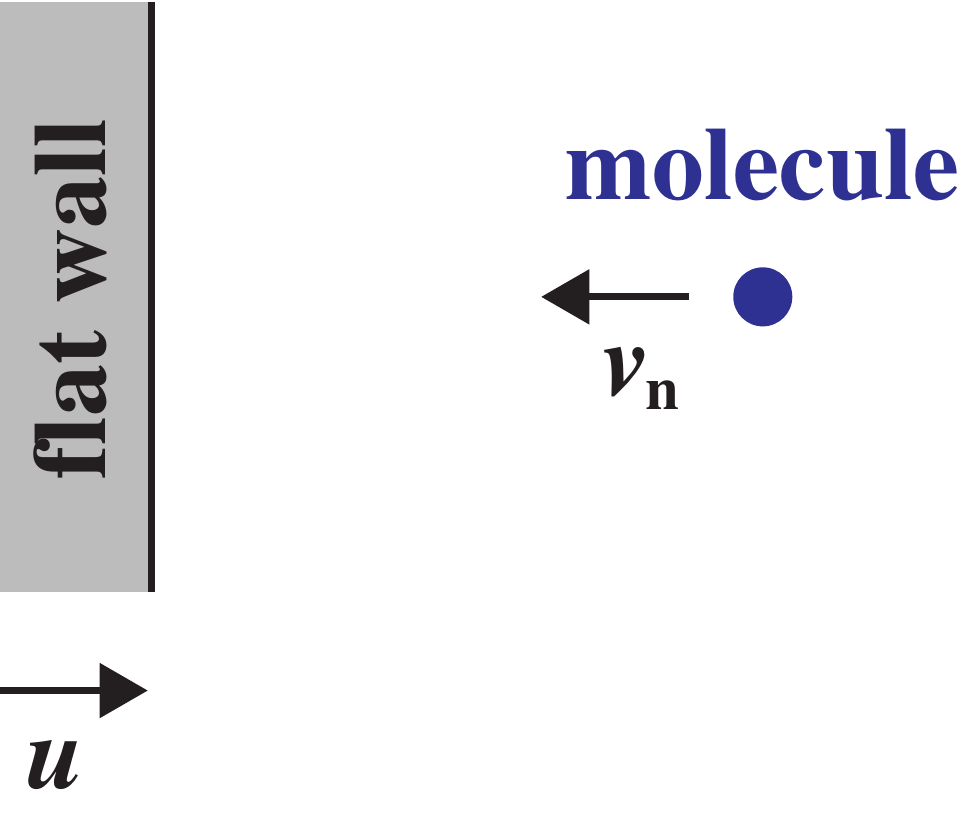}
\end{center}
\caption{
Schematic of a water molecule colliding with a moving flat wall.
The normal component of the relative velocity of the flat wall with respect to the bulk motion of ambient water vapor is $u$, and the normal component of the velocity of water molecules is $v_{\rm n}$.
We set that $u$ is positive when the wall approaches to the water vapor.
}
\label{figx}
\end{figure}

Then we consider the situation that water molecules collide with a moving flat wall (see Figure \ref{figx}).
When the normal component of the relative velocity of the flat wall with respect to the bulk motion of ambient water vapor is $u$, the flux of colliding water molecules is given as a function of $u$ as follows \citep[see also][]{Schrage+1953}:
\begin{eqnarray}
{J_{\rm n} ( u )} & = & \int_{- u}^{+ \infty} n {\left( v_{\rm n} + u \right)} {f ( v_{\rm n} )} ~{\rm d}v_{\rm n} \nonumber \\ 
                  & = & J_{0} {\left[ \exp{\left( - \frac{u^{2}}{{c_{\rm s}}^{2}} \right)} + \sqrt{\pi} \frac{u}{c_{\rm s}} {\left[ \erf{\left( \frac{u}{c_{\rm s}} \right)} + 1 \right]} \right]},
\end{eqnarray}
where $n = P_{\rm water} / {( k_{\rm B} T )}$ is the number density of water molecules, $c_{\rm s} = \sqrt{2 k_{\rm B} T / m}$ is the most probable speed of water molecules, and $\erf$ represents the error function.
Here, we set that $u$ is positive when the wall approaches to the water vapor.

Assuming that chondrules are spherical, the average flux of water molecules, $J$, is given by
\begin{eqnarray}
J & = & \frac{1}{4 \pi} \int_{0}^{\pi} 2 \pi \sin{\theta} {J_{\rm n} ( v_{\rm rel} \cos{\theta})} ~{\rm d}\theta \nonumber \\
  & = & J_{0} {\left( A + B \right)},
\label{eq:J}
\end{eqnarray}
where
\begin{equation}
A = \frac{\sqrt{\pi}}{2} \frac{\erf{\left( s \right)}}{s},
\end{equation}
and
\begin{equation}
B = {\left( s^{2} - \frac{1}{2} \right)} A + \frac{1}{2} \exp{\left( - s^{2} \right)}.
\end{equation}
Here, $s$ is defined as follows:
\begin{equation}
s \equiv \frac{| v_{\rm rel} |}{c_{\rm s}}.
\end{equation}

\begin{figure}
\begin{center}
\includegraphics[width=\columnwidth]{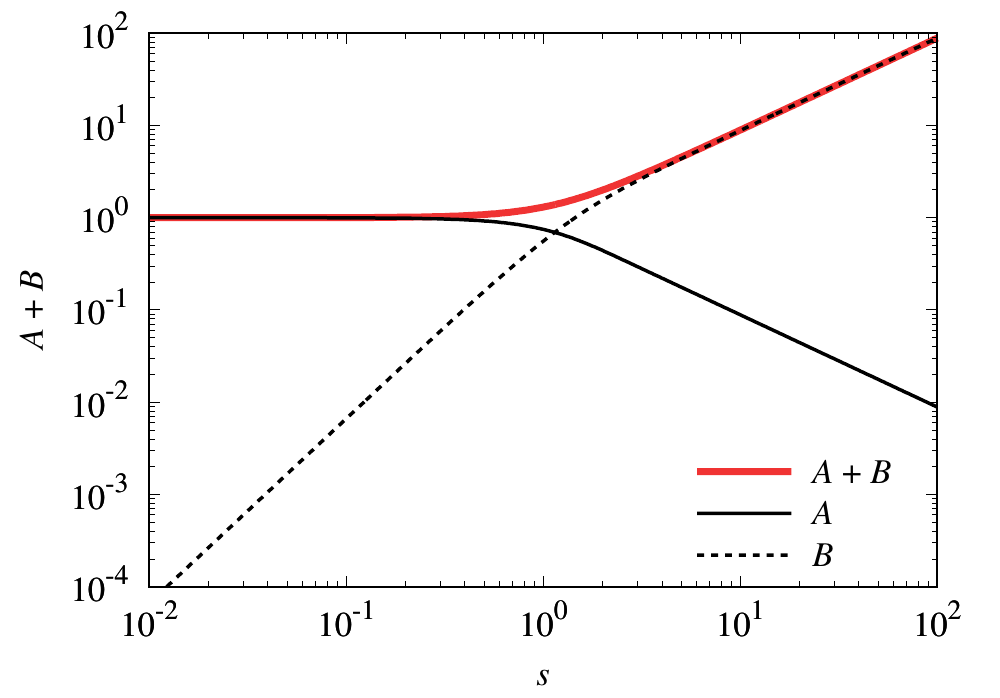}
\end{center}
\caption{
Normalized flux of molecules, $J / J_{0} = A + B$, as a function of $s$.
}
\label{fig1}
\end{figure}

Figure \ref{fig1} shows the normalized flux, $J / J_{0} = A + B$, as a function of $s$.
Here, $A$ and $B$ are the contributions of $c_{\rm s}$ and $v_{\rm rel}$, respectively.
We found that $A = 1 - {( 1 / 3 )} s^{2} + o{(s^{2})}$ and $B = {( 2 / 3 )} s^{2} + o{(s^{2})}$ for the subsonic limit of $s \to 0$, and $A = {( \sqrt{\pi} / 2 )} s^{-1} + o{(s^{-1})}$ and $B = {( \sqrt{\pi} / 2 )} s - {( \sqrt{\pi} / 4 )} s^{-1} + o{(s^{-1})}$ for the supersonic limit of $s \to \infty$.
Thus $J / J_{0} \simeq 1$ for $s \ll 1$ and $J / J_{0} \simeq {( \sqrt{\pi} / 2 )} s$ for $s \gg 1$ as shown in Figure \ref{fig1}.

\begin{figure}
\begin{center}
\includegraphics[width=\columnwidth]{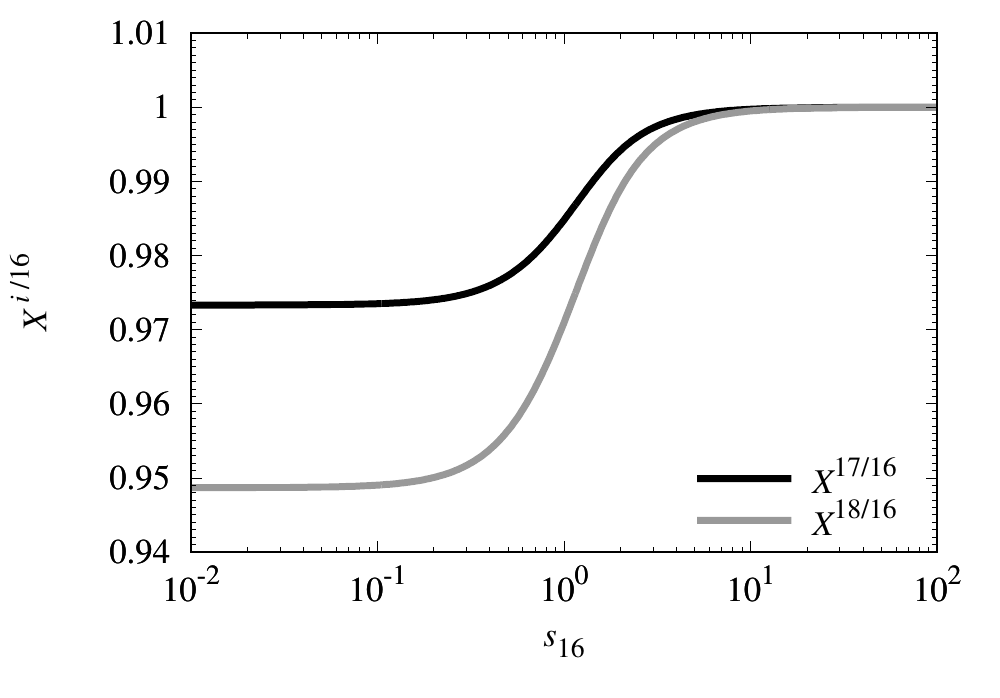}
\end{center}
\caption{
Normalized flux ratio of ${\rm H}_{2} {}^{i}{\rm O}$ and ${\rm H}_{2} {}^{16}{\rm O}$ molecules, $X^{i / 16} = {( J_{i} / J_{16} )} / {( n_{i} / n_{16} )}$, as a function of $s$ for ${\rm H}_{2} {}^{16}{\rm O}$ molecules, $s_{16}$.
}
\label{fig2}
\end{figure}

Oxygen has three isotopes, namely, ${}^{16}{\rm O}$, ${}^{17}{\rm O}$, and ${}^{18}{\rm O}$.
The molecular weight of ${\rm H}_{2} {}^{i}{\rm O}$, $m_{i}$ $(i = 16, 17, 18)$, is approximately given by $m_{i} = {( i + 2 )} m_{\rm H}$, where $m_{\rm H}$ is the mass of hydrogen atoms.\footnote{In this study, we do not consider the presence of deuterium (D) in water molecules for simplicity.}
The value of $c_{\rm s}$ differs among different isotopes, and $s$ depends on the isotope species $i$.
When the number density of ${\rm H}_{2} {}^{i}{\rm O}$ molecules in the ambient vapor is $n_{i}$, the flux ratio of ${\rm H}_{2} {}^{i}{\rm O}$ and ${\rm H}_{2} {}^{j}{\rm O}$ molecules, $J_{i} / J_{j}$, is not equal to the number density ratio, $n_{i} / n_{j}$.
Figure \ref{fig2} shows the normalized flux ratio of ${\rm H}_{2} {}^{i}{\rm O}$ and ${\rm H}_{2} {}^{16}{\rm O}$ molecules, 
\begin{equation}
X^{i / 16} = \frac{J_{i} / J_{16}}{n_{i} / n_{16}}, 
\end{equation}
as a function of $s$ for ${\rm H}_{2} {}^{16}{\rm O}$ molecules, $s_{16}$.
The dependence of $s_{16}$ on $T$ and $v_{\rm rel}$ is as follows:
\begin{equation}
s_{16} = 0.74 {\left( \frac{T}{2000~{\rm K}} \right)}^{-1/2} {\biggl( \frac{v_{\rm rel}}{1~{\rm km}~{\rm s}^{-1}} \biggr)}.
\end{equation}
We found that $X^{i / 16} = \sqrt{ m_{16} / m_{i}}$ when $s_{16} = 0$, and $X^{i / 16} \to 1$ in the hypersonic limit of $s_{16} \to \infty$.

It is important to note that not all molecules colliding with chondrule surface can cause oxygen isotope exchange reaction.
The influx of water molecules which can react with chondrules, $J_{i}^{\rm in}$, is given by
\begin{equation}
J_{i}^{\rm in} = \beta_{i} J_{i},
\label{eq:beta}
\end{equation}
where $\beta_{i}$ is the isotope exchange efficiency of colliding ${\rm H}_{2} {}^{i}{\rm O}$ molecules \citep[e.g.,][]{2018ApJ...865...98Y, 2021GeCoA.314..108Y}.

\subsection{Efflux of water molecules} \label{sec:efflux}

In this section, we derive theoretical constraints on the efflux of water molecules (per unit area) at the chondrule surface from the point of view of isotope fractionation factors at equilibrium.
We introduce the isotope ratio of the water vapor, $R^{i / 16}_{\rm vapor} = n_{i} / n_{16}$, and the isotope ratio of chondrules, $R^{i / 16}_{\rm chondrule}$, for $i = 17$ and $18$.

Here, we consider the case that the isotope exchange reaction is in steady state and $v_{\rm rel} = 0$.
At the equilibrium, the ratio of $R^{i / 16}_{\rm chondrule}$ to $R^{i / 16}_{\rm vapor}$ is equal to the isotope fractionation factor, $\alpha^{i / 16}$;
\begin{equation}
\frac{R^{i / 16}_{\rm chondrule}}{R^{i / 16}_{\rm vapor}} = \alpha^{i / 16}.
\end{equation}
The efflux of ${\rm H}_{2} {}^{i}{\rm O}$ molecules after isotope exchange reaction, $J_{i}^{\rm out}$, should be balanced with the influx:
\begin{equation}
J_{i}^{\rm out} = J_{0, i}^{\rm in},
\end{equation}
where $J_{0, i}^{\rm in}$ represents $J_{i}^{\rm in}$ for $v_{\rm rel} = 0$.
We can rewrite the above equation as
\begin{eqnarray}
\frac{J_{i}^{\rm out}}{J_{16}^{\rm out}} & = & \frac{J_{0, i}^{\rm in}}{J_{0, 16}^{\rm in}} \nonumber \\
                                         & = & \frac{\beta_{i}}{\beta_{16}} \sqrt{\frac{m_{16}}{m_{i}}} R^{i / 16}_{\rm vapor} \nonumber \\
                                         & = & \frac{1}{\alpha^{i / 16}} \frac{\beta_{i}}{\beta_{16}} \sqrt{\frac{m_{16}}{m_{i}}} R^{i / 16}_{\rm chondrule}.
\label{eq:efflux_i/16}
\end{eqnarray}

Equation (\ref{eq:efflux_i/16}) shows the general expression of $J_{i}^{\rm out} / J_{16}^{\rm out}$.
It is known that the isotope fractionation factor depends on the temperature, and its deviation from $1$ is sufficiently small ($\lesssim 1\tcperthousand$) when the temperature exceeds the melting point of chondrules \citep[e.g.,][]{stolper1991experimental, 2003GeCoA..67..459A}.
Laboratory experiments on the isotope exchange reaction between silicate melt and water vapor \citep[e.g.,][]{2015GeCoA.164...17D} also indicate that $\alpha^{i / 16}$ would be close to $1$ around the melting point.
It is also expected that the dependence of $\beta_{i}$ on $i$ is negligibly small \citep[i.e., $\beta_{i} / \beta_{16} \simeq 1$;][]{2021GeCoA.314..108Y}.
Therefore, we assume that $J_{i}^{\rm out} / J_{16}^{\rm out}$ is given by the following equation:
\begin{equation}
\frac{J_{i}^{\rm out}}{J_{16}^{\rm out}} = \sqrt{\frac{m_{16}}{m_{i}}} R^{i / 16}_{\rm chondrule}.
\label{eq:efflux_i/16_mod}
\end{equation}

In this study, we assume that the efflux at the chondrule surface follows Equation (\ref{eq:efflux_i/16_mod}) not only for the case of $v_{\rm rel} = 0$ but also for $v_{\rm rel} \neq 0$.
Although whether this assumption is correct or not is experimentally unclear, it would be natural when the efflux is controlled by the activation energies for isotope exchange reaction at surface, which depends on the isotope species $i$ but not on $v_{\rm rel}$.

The conservation of the number of oxygen atoms in chondrules is also required for isotope exchange reactions.\footnote{
We note that the conservation does not hold when evaporation or condensation of silicate melt occurs (see Section \ref{sec:evaporation}).
In addition, when iron metals in chondrules oxidize, the number of oxygen atoms in chondrules also increases (see Section \ref{sec:oxidization}).
Indeed, some of chondrules would have experienced net evaporation and recondensation during olivine crystallization \citep[e.g.,][]{2018SciA....4.3321L, 2019PNAS..11623461M}.}
The equation for the conservation is given by
\begin{equation}
\sum_{i} J_{i}^{\rm out} = \sum_{i} J_{i}^{\rm in} = J_{\rm sum}.
\end{equation}

\subsection{Temporal evolution of oxygen isotope ratio}

The oxygen isotope ratio of chondrules is calculated from the number of oxygen atoms in a chondrule.
The total number of oxygen atoms in a chondrule, $N_{\rm sum}$, is the sum of ${}^{i}{\rm O}$ atoms in a chondrule, $N_{i}$:
\begin{equation}
N_{\rm sum} = \sum_{i} N_{i}.
\end{equation}
The temporal evolution of $N_{i}$ is given by the following equation:
\begin{equation}
\frac{{\rm d}N_{i}}{{\rm d}t} = 4 \pi r^{2} {\left( J_{i}^{\rm in} - J_{i}^{\rm out} \right)},
\label{eq:dNdt}
\end{equation}
where $t$ is the time and $r$ is the chondrule radius.
Here, we assume that all chondrule surface can react with ambient water vapor and the oxygen isotope ratio is homogeneous within the chondrule for simplicity.
The isotope ratio of chondrule is given by
\begin{equation}
R^{i / 16}_{\rm chondrule} = \frac{N_{i}}{N_{16}}.
\end{equation}

The total number of oxygen atoms in a chondrule depends on its composition.
Here, we consider the situation that the composition of chondrules is expressed by a single chemical formula for simplicity.
When the molar volume is $\Omega$ and the number of oxygen atoms in the chemical formula is $Z_{\rm oxy}$, $N_{\rm sum}$ is given by
\begin{equation}
N_{\rm sum} = \frac{4 \pi r^{3} N_{\rm A}}{3 \Omega} Z_{\rm oxy}, 
\end{equation}
where $N_{\rm A}$ is Avogadro’s number.
In this study, we assume that chondrules are made of pure molten forsterite (${\rm Mg}_{2}{\rm Si}{\rm O}_{4}$) whose molar volume is $\Omega = 50~{\rm cm}^{3}~{\rm mol}^{-1}$ and number of oxygen atoms in the chemical formula is $Z_{\rm oxy} = 4$.
For simplicity, we assume that $\Omega$ and $Z_{\rm oxy}$ are constant over time.

\subsection{The $\delta$ notation}

Here, we introduce the $\delta$ notation to measure the derivation of isotope ratio from the terrestrial reference material.
Using the $\delta$ notation, the isotope ratio of a sample (i.e., chondrule or vapor), ${\delta{}^{i}{\rm O}}_{\rm sample}$, is defined as follows:
\begin{equation}
{\delta{}^{i}{\rm O}}_{\rm sample} = {\left( \frac{R^{i / 16}_{\rm sample}}{R^{i / 16}_{\rm VSMOW}} - 1 \right)} \times 10^{3} \tcperthousand, 
\end{equation}
where $R^{i / 16}_{\rm sample}$ is the oxygen isotope ratio of sample and $R^{i / 16}_{\rm VSMOW}$ is the oxygen isotope ratio of the terrestrial reference material called Vienna Standard Mean Ocean Water (VSMOW).
The oxygen isotope ratios of VSMOW are $R^{17 / 16}_{\rm VSMOW} = 379.9 \times 10^{-6}$ and $R^{18 / 16}_{\rm VSMOW} = 2005.2 \times 10^{-6}$.

\section{Results}
\label{sec:results}

In this section, we show the results of some example calculations to demonstrate the impacts of nonzero relative velocity on the oxygen isotope exchange reaction of chondrules.

\subsection{Oxygen isotope ratio of chondrules in steady state}

First, we calculate the oxygen isotope ratio of chondrules in steady state.
We discuss the ``effective'' fractionation factor and its dependence on $s_{16}$ in Section \ref{sec:mass-dependent}.
We also calculate the oxygen isotope ratio of chondrules in the situation where the chondrules and water vapor are in a closed system in Section \ref{sec:balance}.

\subsubsection{Mass-dependent fractionation caused by isotope exchange reaction}
\label{sec:mass-dependent}

\begin{figure}
\begin{center}
\includegraphics[width=\columnwidth]{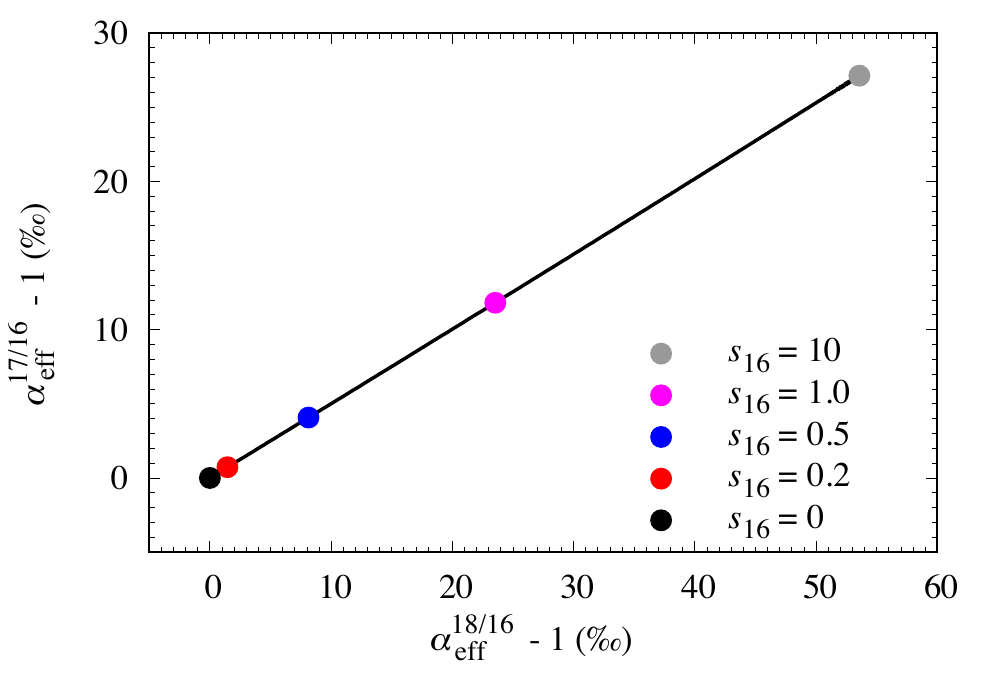}
\end{center}
\caption{
Deviation of $\alpha^{i / 16}_{\rm eff}$ from $1$ (i.e., $\alpha^{i / 16}_{\rm eff} - 1$) in the style of the oxygen three-isotope plot.
}
\label{fig4}
\end{figure}

We can calculate the oxygen isotope ratio of chondrules in steady state by solving Equation (\ref{eq:dNdt}) with ${{\rm d}N_{i}} / {{\rm d}t} = 0$ for all $i$.
We found that $R^{i / 16}_{\rm chondrule}$ (for $i = 17$ and $18$) is given by
\begin{equation}
R^{i / 16}_{\rm chondrule} = \sqrt{\frac{m_{i}}{m_{16}}} X^{i / 16} R^{i / 16}_{\rm vapor}.
\end{equation}
We introduce the ``effective'' fractionation factor,
\begin{eqnarray}
\alpha^{i / 16}_{\rm eff} & \equiv & \frac{R^{i / 16}_{\rm chondrule}}{R^{i / 16}_{\rm vapor}} \nonumber \\
                         & = & \sqrt{\frac{m_{i}}{m_{16}}} X^{i / 16}.
\label{eq:balance}
\end{eqnarray} 
Figure \ref{fig4} shows the deviation of $\alpha^{i / 16}_{\rm eff}$ from $1$ as a function of $s_{16}$.
For $v_{\rm rel} = 0$, it is clear that the following equation,
\begin{equation}
\alpha^{i/ 16}_{\rm eff} = 1,
\end{equation}
is satisfied in steady state.
In contrast, we found that $\alpha^{i / 16}_{\rm eff}$ deviates from $1$ when $v_{\rm rel} \neq 0$.
For the supersonic limit, $\alpha^{i / 16}_{\rm eff}$ becomes
\begin{equation}
\alpha^{i / 16}_{\rm eff} = \sqrt{\frac{m_{i}}{m_{16}}},
\end{equation}
and mass-dependent isotope fractionation occurs in this case.
The deviation of $\alpha^{i / 16}_{\rm eff}$ from $1$ is approximately $27 \tcperthousand$ for $i = 17$ and $54 \tcperthousand$ for $i = 18$ when $s_{16} \to \infty$.

\subsubsection{Deviation from the mixing line}
\label{sec:balance}

We also calculate the oxygen isotope ratio of chondrules in the situation where the oxygen isotope ratio of water vapor is also altered via isotope exchange reaction.
Here, we assume that oxygen atoms exist only in water molecules, and we do not consider other oxides including carbon monoxide (${\rm C}{\rm O}$) and carbon dioxide (${\rm C}{\rm O}_{2}$).
When the number density of chondrules in space, $n_{\rm chondrule}$, is constant over time, the following equation for the conservation,
\begin{equation}
N_{i} n_{\rm chondrule} + n_{i} = {\rm const.},
\label{eq:cons}
\end{equation}
is satisfied for all $i$.
We define the chondrule-to-vapor oxygen molar ratio, $\chi$, as
\begin{equation}
\chi = \frac{N_{\rm sum} n_{\rm chondrule}}{n_{\rm sum}},
\end{equation}
where
\begin{equation}
n_{\rm sum} = \sum_{i} n_{i}
\end{equation}
is the total number density of water molecules in the ambient vapor.
Assuming that the dust-to-ice mass ratio of the solar nebula is on the order of $1$ \citep[e.g.,][]{1981PThPS..70...35H, 2018ApJ...869L..45B} and the majority of dust is in the form of chondrules while water is in vapor phase, we obtain $\chi \sim 1$ as a reference value.

\begin{figure}
\begin{center}
\includegraphics[width=\columnwidth]{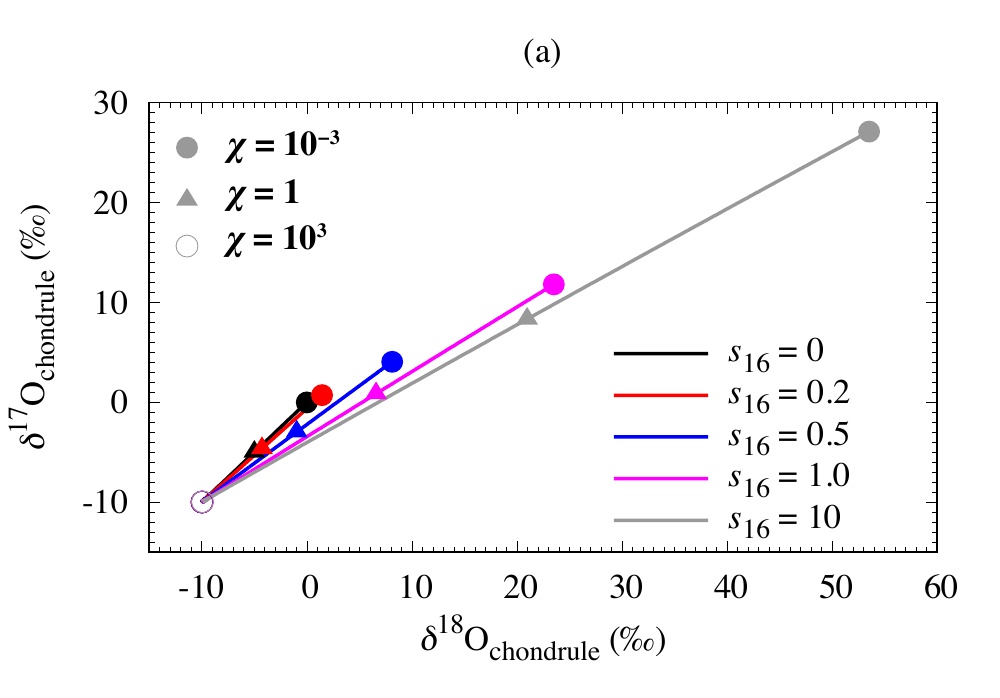}
\includegraphics[width=\columnwidth]{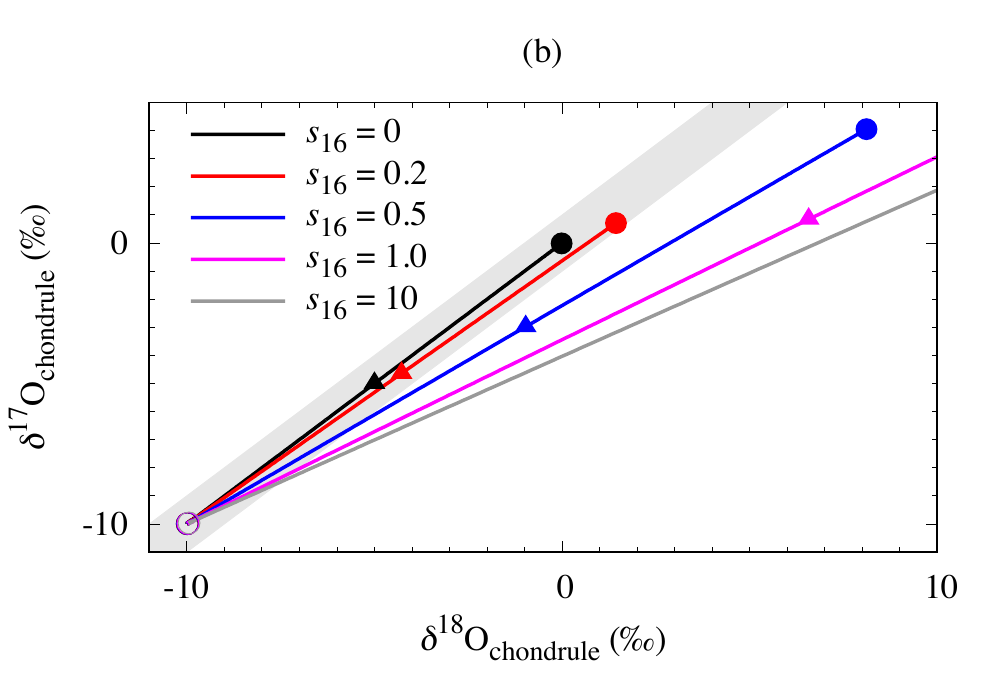}
\end{center}
\caption{
(a) Oxygen three-isotope plot for chondrules in steady state as a function of $s_{16}$.
Here, we assume that, at $t = 0$, the oxygen isotope ratios of chondrules and water vapor are ${\delta{}^{17}{\rm O}}_{\rm chondrule} = {\delta{}^{18}{\rm O}}_{\rm chondrule} = - 10 \tcperthousand$ and ${\delta{}^{17}{\rm O}}_{\rm vapor} = {\delta{}^{18}{\rm O}}_{\rm vapor} = 0 \tcperthousand$, respectively.
Filled circles represent results for $\chi = 10^{-3}$, filled triangles are for $\chi = 1$, and open circles are for $\chi = 10^{3}$.
(b) Close-up view of Panel (a).
The gray shaded region denotes the area that the deviation of ${\delta{}^{17}{\rm O}}_{\rm chondrule}$ from the mixing line (${\delta{}^{17}{\rm O}}_{\rm chondrule} = {\delta{}^{18}{\rm O}}_{\rm chondrule}$) is less than $1 \tcperthousand$.
}
\label{fig5}
\end{figure}

Figure \ref{fig5} shows the oxygen isotope ratios of chondrules in steady state as a function of $s_{16}$ and $\chi$.
Here, we assume that, at $t = 0$, the oxygen isotope ratios of chondrules and water vapor are ${\delta{}^{17}{\rm O}}_{\rm chondrule} = {\delta{}^{18}{\rm O}}_{\rm chondrule} = - 10 \tcperthousand$ and ${\delta{}^{17}{\rm O}}_{\rm vapor} = {\delta{}^{18}{\rm O}}_{\rm vapor} = 0 \tcperthousand$, respectively.
The mixing line is therefore ${\delta{}^{17}{\rm O}} = {\delta{}^{18}{\rm O}}$, and the evolution of the oxygen isotope ratio of chondrule melt will follow this mixing line when $s_{16} = 0$ (see gray shaded region in Figure \ref{fig5}(b)).

We found that the oxygen isotope ratios of chondrules deviates from the mixing line when $s_{16} \ne 0$.
When $s_{16} \gtrsim 0.2$, the deviation of the oxygen isotope ratios from the mixing line is larger than $1\tcperthousand$ in ${\delta{}^{18}{\rm O}}_{\rm chondrule}$ value.
As the deviation observed in chondrules in primitive carbonaceous chondrites is within a few permil \citep[e.g.,][]{2018crpd.book..192T, 2022ChGeo.608l1016Z}, our results indicate that these chondrules formed under the situation of $s_{16} \lesssim 0.2$.

We note that the oxygen isotope ratio of chondrules barely depends on $s_{16}$ when $\chi \gg 1$.
This feature is understandable from the mass balance in a closed system.
For $\chi \gg 1$, most of oxygen atoms are in chondrules, and the average isotope ratio of chondrules and gas is approximately equal to the isotope ratio of chondrules.
Therefore, the oxygen isotope ratio of chondrules barely changes while that of the residual gas is significantly enriched in $^{16}{\rm O}$.
The degree of isotopic fractionation in the gas is determined by Equation (\ref{eq:balance}).

\begin{figure}
\begin{center}
\includegraphics[width=\columnwidth]{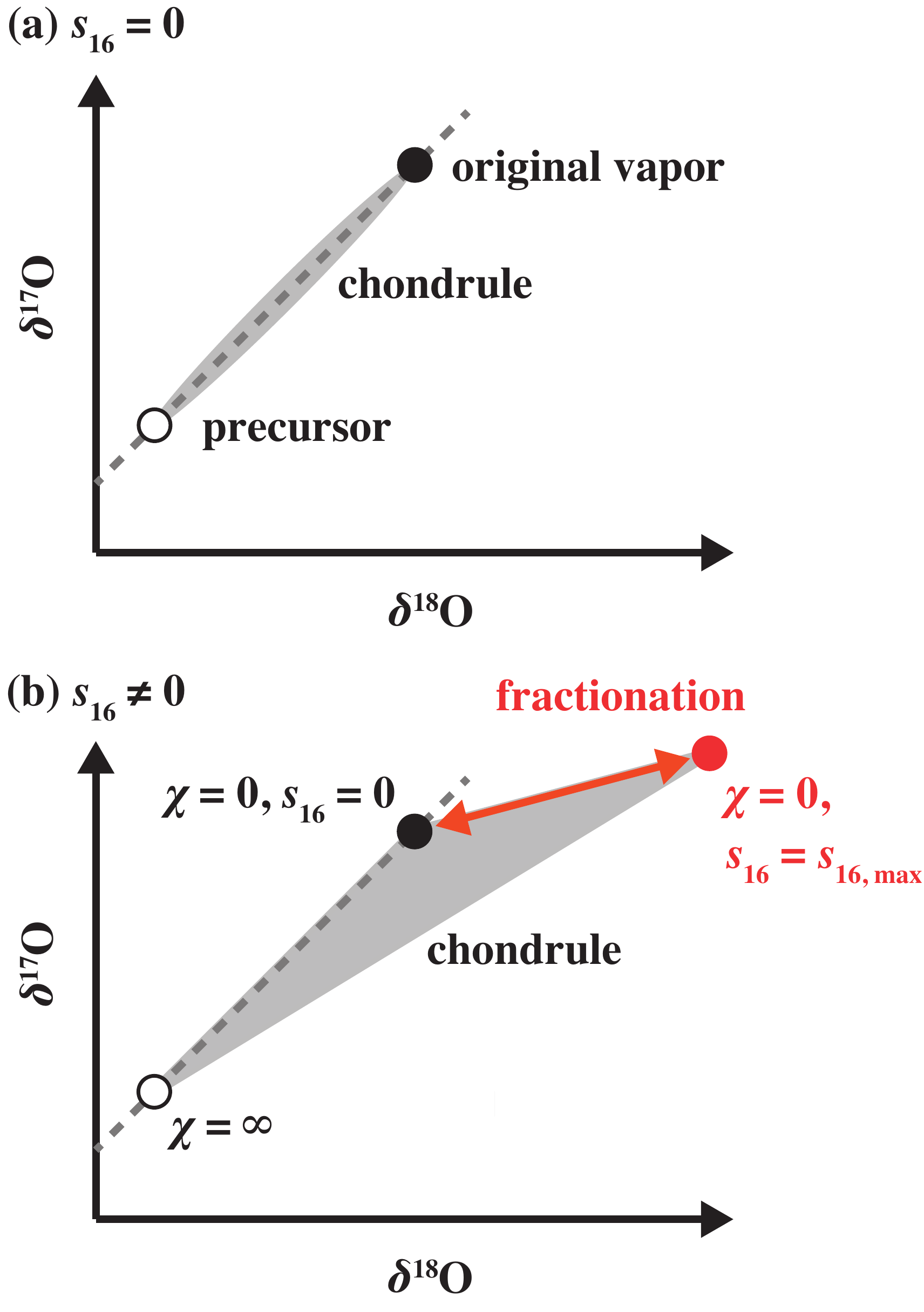}
\end{center}
\caption{
Schematic of the variation of oxygen isotope ratios of chondrules as a results of mass-dependent fractionation caused by nonzero relative velocity.
The black circle denotes the oxygen isotope ratios of original ambient vapor before the isotope exchange reaction, and the white is the oxygen isotope ratios of chondrule precursors before the isotope exchange reaction.
(a) When $s_{16} = 0$, the oxygen isotope ratios of chondrules are plotted along the mixing line (gray dashed line), which would be identical to the PCM line.
(b) When $s_{16} \neq 0$, the oxygen isotope ratios of chondrules are plotted in the shaded triangle region.
We assume that each chondrule was formed under a condition with different values of $s_{16}$ and $\chi$.
The orange circle denotes the oxygen isotope ratios of chondrules after the isotope exchange reaction for the maximum value of $s_{16}$ in the formation environment ($s_{16, {\rm max}}$) and the limit of $\chi \ll 1$ (i.e., filled circles in Figure \ref{fig5}).
The width of the shaded triangle (orange double arrow) increases with the maximum value of $s_{16}$.
}
\label{fig7}
\end{figure}

Figure \ref{fig7} shows the schematic of the variation of oxygen isotope ratios of chondrules after the isotope exchange reaction.
Panel (a) shows the result for $s_{16} = 0$, and panel (b) shows the result for $s_{16} \neq 0$.
Here, we assume that the oxygen isotope ratios of chondrules are determined by the mass balance of chondrule precursors and the ambient water vapor (Equation (\ref{eq:cons})), and we do not consider the effects of other oxides for simplicity.

It is clear that the final oxygen isotope ratios of chondrules are on the mixing line of the compositions of chondrule precursors and ambient vapor (i.e., PCM line) when the relative velocity between molten chondrules and ambient vapor is $0$ (i.e., $s_{16} = 0$).
In contrast, if $s_{16} \neq 0$, the final oxygen isotope ratios of chondrules deviate from the PCM line, and the deviation depends on $s_{16}$.
Larger chondrules would have larger $s_{16}$ due to their inertia, and $s_{16}$ significantly varies among chondrules as they have size distribution.
Therefore, the final oxygen isotope ratios of chondrules would not be on a single line; instead, oxygen isotope ratios would be plotted in a triangle region as illustrated in Figure \ref{fig7}(b).
This is, however, inconsistent with the observed variation in carbonaceous chondrites \citep[e.g.,][]{2018crpd.book..192T, 2022ChGeo.608l1016Z}.

\subsection{Timescale for isotope exchange reaction}
\label{sec:tau}

Then we evaluate the timescale for isotope exchange reaction.
Equation (\ref{eq:dNdt}) shows the time derivative of $N_{i}$.
When the oxygen isotope ratio of water vapor is constant over time, the timescale for isotope exchange reaction, $\tau$, is given by
\begin{eqnarray}
\tau & = & \frac{N_{\rm sum}}{4 \pi r^{2} J_{\rm sum}} \nonumber \\
     & = & \frac{Z_{\rm oxy} N_{\rm A} r}{3 \Omega J_{\rm sum}}.
\end{eqnarray}
If we consider the temporal change of the oxygen isotope ratio of water vapor, the timescale should be reduced.
Based on the conservation equation (Equation (\ref{eq:cons})), the time differential for ambient vapor, ${{\rm d}{( n_{i} / n_{\rm sum} )}} / {{\rm d}t}$, is proportional to that for chondrules, ${{\rm d}{( N_{i} / N_{\rm sum} )}} / {{\rm d}t}$, and we obtain the following equation:
\begin{equation}
\frac{{\rm d}{( N_{i} / N_{\rm sum}} - n_{i} / n_{\rm sum} )}{{\rm d}t} = {\left( 1 + \chi \right)} \frac{{\rm d}{( N_{i} / N_{\rm sum} )}}{{\rm d}t}.
\end{equation}
Therefore, $\tau$ is reduced by a factor of ${( 1 + \chi )}$ in this case;
\begin{equation}
\tau = \frac{Z_{\rm oxy} N_{\rm A} r}{3 {( 1 + \chi )} \Omega J_{\rm sum}}.
\label{eq:tau_chi}
\end{equation}

\begin{figure}
\begin{center}
\includegraphics[width=\columnwidth]{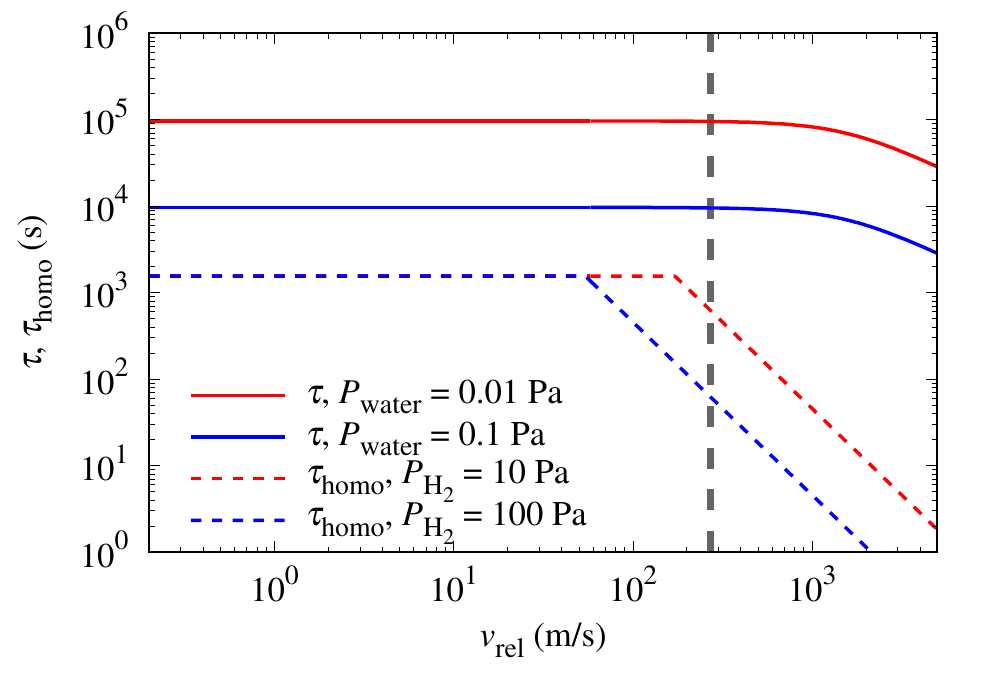}
\end{center}
\caption{
Timescale for isotope exchange reaction, $\tau$, as a function of $P_{\rm water}$ and $v_{\rm rel}$ (solid lines).
The vertical dashed line represents $v_{\rm rel}$ at $s_{16} = 0.2$ (see Section \ref{sec:balance}).
Here, we set $T = 2000~\si{K}$, $r = 0.5~\si{mm}$, $\chi = 1$, and $\beta_{16} = 0.3$.
The dashed lines represent the timescale of homogenization, $\tau_{\rm homo}$, as a function of $P_{{\rm H}_{2}}$ and $v_{\rm rel}$ (see Section \ref{sec:homo}).
}
\label{fig6}
\end{figure}

Figure \ref{fig6} shows the timescale for isotope exchange reaction, $\tau$, as a function of ${\rm H}_{2}{\rm O}$ partial pressure, $P_{\rm water}$, and $v_{\rm rel}$.
Here, we assume $J_{\rm sum} = \beta_{16} J_{16}$ and $P_{\rm water} = n_{16} k_{\rm B} T$ for simplicity.
We also set $T = 2000~\si{K}$, $r = 0.5~\si{mm}$, $\chi = 1$, and $\beta_{16} = 0.3$ in Figure \ref{fig6}.
We found that $\tau$ barely depends on $v_{\rm rel}$ for $v_{\rm rel} \ll 1~\si{km}~\si{s}^{-1}$, and $\tau$ is inversely proportional to $P_{\rm water}$.
For $v_{\rm rel} = 0$, the timescale for isotope exchange reaction defined by Equation (\ref{eq:tau_chi}) is approximately given by
\begin{eqnarray}
\tau & \simeq & \frac{Z_{\rm oxy} N_{\rm A} r}{3 {( 1 + \chi )} \Omega \beta_{16} J_{16}} \nonumber \\
     & \simeq & 9.7 \times 10^{3} \cdot {\left( \frac{1 + \chi}{2} \right)}^{-1} {\left( \frac{\beta_{16}}{0.3} \right)}^{-1} \nonumber \\
     &        & \cdot {\left( \frac{P_{\rm water}}{0.1~\si{Pa}} \right)}^{-1} {\left( \frac{T}{2000~\si{K}} \right)}^{1/2} {\biggl( \frac{r}{0.5~\si{mm}} \biggr)}~\si{s}.
\label{eq:tau}
\end{eqnarray}

The fraction of oxygen atoms in a chondrule which is exchanged with ambient vapor, $f {( t )}$, is given as follows; 
\begin{equation}
f {( t )} = 1 - \exp{\left( - \frac{t}{\tau} \right)},
\end{equation}
and approximately 90\% of oxygen atoms in the chondrule experience isotope exchange at $t \simeq 2 \tau$.

\section{Discussion}
\label{sec:discussion}

Oxygen isotope compositions of chondrules in carbonaceous chondrites have been studied by several groups \citep[e.g.,][]{1984E&PSL..67..151C, 2010GeCoA..74.2473C, 2017M&PS...52.2672J, 2018GeCoA.224..116H, 2018E&PSL.496..132M, 2022GeCoA.322..194F}, and it is widely accepted that the oxygen isotope compositions scatter along the PCM line on the three oxygen isotope plot \citep{2012GeCoA..90..242U}.
The deviation of ${\delta{}^{17}{\rm O}}$ of each chondrule from the PCM line is typically within $\pm 1 \tcperthousand$.
In this section, we briefly discuss the formation environment of chondrules from the point of view of oxygen isotopic compositions.

\subsection{Dynamics of chondrules in post-shock region}
\label{sec:shock}

The leading candidate of chondrule-forming shock is a bow shock caused by planetesimals/protoplanets with high heliocentric eccentricity.
Here, we carry out an order-of-magnitude estimate on the dynamics of chondrules in post-shock region, and we discuss whether bow shocks are plausible for the origin of chondrules or not.

In the post-shock region, the dynamics of chondrules approaches to the quasi-steady state when the distance from the shock front exceeds the stopping length of chondrules owing to gas drag.
\citet{2022ApJ...927..188A} found that $v_{\rm rel}$ is approximately given by
\begin{equation}
v_{\rm rel} \simeq 0.64 \frac{\rho}{\rho_{\rm gas}} \frac{v_{\rm gas}}{c_{\rm gas}} \frac{| v_{\rm gas} - v_{0} |}{L} r,
\label{eq:vrel}
\end{equation}
where $\rho$ is the material density of chondrules, $\rho_{\rm gas}$ is gas density, $v_{\rm gas}$ is the post-shock gas velocity with respect to the shock front, $c_{\rm gas}$ is the most probable speed of the nebular gas molecules in the post-shock region, $v_{0}$ is the pre-shock gas velocity, and $L$ is the length scale of the shock (see \ref{app:vrel}).\footnote{
We can roughly evaluated the duration of melting of chondrules as $\tau_{\rm melt} \sim L / v_{\rm gas}$.
When we consider bow shocks caused by planetesimals, $L < 10^{4}~\si{km}$ and $v_{\rm gas} \sim 1~\si{km.s^{-1}}$ in the post-shock region, and we obtain $\tau_{\rm melt} < 10^{4}~\si{s}$.
This upper limit is comparable with the timescale for isotope exchange reaction when $P_{\rm water} \sim 10^{-1}~\si{Pa}$ (see Figure \ref{fig6}).
Therefore the oxygen isotope exchange reaction could occur efficiently when $P_{\rm water} \gg 10^{-1}~\si{Pa}$ in the post-shock region.
However, $v_{\rm rel}$ in the post-shock region would be too large to reproduce the small mass-dependent fractionation observed in chondrules.
}
Assuming that the nebular gas is ${\rm H}_{2}$ gas whose temperature is $2000~{\rm K}$, $v_{0} = 10~{\rm km}~{\rm s}^{-1}$, $v_{\rm gas} = 0.2 v_{0}$, and $\rho = 3~{\rm g}~{\rm cm}^{-3}$, we can rewrite Equation (\ref{eq:vrel}) as follows:
\begin{eqnarray}
v_{\rm rel} & \simeq & 4 \times 10^{2} \cdot {\left( \frac{\rho_{\rm gas}}{10^{-9}~{\rm g}~{\rm cm}^{-3}} \right)}^{-1} \nonumber \\
            &        & \cdot {\left( \frac{L}{10^{4}~{\rm km}} \right)}^{-1} {\biggl( \frac{r}{0.5~{\rm mm}} \biggr)}~{\rm m}~{\rm s}^{-1}.
\end{eqnarray}

We note that $v_{\rm rel} \simeq 4 \times 10^{2} {( r / 0.5~{\rm mm} )}~{\rm m}~{\rm s}^{-1}$ is the lower bound estimate, and $v_{\rm rel}$ in the chondrule formation region might be much higher than that value.
Based on the minimum mass solar nebula model \citep{1981PThPS..70...35H}, the gas density at the disk midplane is
\begin{equation}
\rho_{\rm gas, mid} = 1.4 \times 10^{-9} {\left( \frac{R}{1~{\rm au}} \right)}^{- 11/4}~{\rm g}~{\rm cm}^{-3},
\end{equation}
where $R$ is the heliocentric distance.
When we consider the excitations of planetesimals caused by the resonances of Jupiter in the gaseous solar nebula as a source of shock waves, strong bow shocks whose $v_{0}$ is sufficiently large to make chondrules occur at $R \sim 2~{\rm au}$ \citep[e.g.,][]{2014ApJ...794L...7N, 2019ApJ...871..110N}.
Once gas accretion onto Jupiter starts, the nebular gas density inside the Jupiter's orbit decreases significantly \citep[e.g.,][]{2016ApJ...823...48T, 2020ApJ...891..143T}.
Therefore, we can expect that $\rho_{\rm gas}$ in the pre-shock region would be lower than $10^{-10}~{\rm g}~{\rm cm}^{-3}$, and $\rho_{\rm gas}$ in the post-shock region would be lower than $10^{-9}~{\rm g}~{\rm cm}^{-3}$ even if we take it into consideration that $\rho_{\rm gas}$ is enhanced approximately one order of magnitude when the gas passes through the shock front \citep[e.g.,][]{2001Icar..153..430I, 2018crpd.book..375M}.\footnote{
\citet{2007ApJ...671..878D} proposed an alternative mass distribution model for the initial solar nebula.
The model predicts that the gas density in the pre-shock region is $\rho_{\rm gas} \sim 3 \times 10^{-9}~{\rm g}~{\rm cm}^{-3}$ at $R \sim 2~\si{au}$, and $\rho_{\rm gas}$ is higher than $10^{-8}~{\rm g}~{\rm cm}^{-3}$ in the post-shock region.
In this case, $v_{\rm rel}$ becomes sufficiently lower than $10^{2}~\si{m.s^{-1}}$ and the isotopic fractionations would be negligibly small.
}
Moreover, if chondrules in carbonaceous chondrites formed beyond the current Jupiter's orbit ($5.2~{\rm au}$ from the Sun), $\rho_{\rm gas}$ would be orders of magnitude lower than $10^{-9}~{\rm g}~{\rm cm}^{-3}$.

The length scale of the shock, $L$, would be a few to several times larger than the radius of excited planetesimals \citep[e.g.,][]{2010ApJ...719..642M, 2012ApJ...752...27M}.
The biggest asteroid in the main belt is (1) Ceres, and its radius is approximately $500~{\rm km}$.
If the radius of excited planetesimals is not larger than $10^{3}~{\rm km}$, then $L$ might be smaller than $10^{4}~{\rm km}$.

In conclusion, it would be expected that $v_{\rm rel}$ would be significantly higher than $100~{\rm m}~{\rm s}^{-1}$ for chondrules with $r = 0.5~{\rm mm}$ when chondrules formed via shock waves around excited planetesimals in the gaseous solar nebula.
As the oxygen isotope ratio of chondrules in carbonaceous chondrites indicates that $v_{\rm rel}$ is lower than approximately $300~{\rm m}~{\rm s}^{-1}$ (i.e., $s_{16} \lesssim 0.2$; see Figure \ref{fig5}), the shock wave model might not be plausible for the origin of chondrules in carbonaceous chondrites.\footnote{
We acknowledge, however, that the constraint on $v_{\rm rel}$ could be mitigated when we consider additional effects including evaporation/recondensation and oxidization/reduction (see Sections \ref{sec:evaporation} and \ref{sec:oxidization}).
Future studies on these points would be essential.}${}^{,}$\footnote{
\citet{2014ApJ...797...30J} noted that planetesimal bow shocks would also be problematic in the context of collisional destruction of chondrules.
}

\subsection{Timescale of isotopic homogenization within a chondrule}
\label{sec:homo}

As the oxygen isotope composition within a chondrule is homogeneous except for relict minerals \citep[e.g.,][]{2012GeCoA..90..242U}, the duration of melting of chondrules would be longer than the timescale of isotopic homogenization within a chondrule.
Here, we consider the two mechanisms for isotopic homogenization: isotope diffusion within a melt and mechanical mixing due to forced convection.
The timescale of isotopic homogenization should be given as the shorter one of them.

The timescale of isotope diffusion in a chondrule is given by
\begin{equation}
\tau_{\rm diff} = \frac{r^{2}}{D},
\end{equation}
where $D$ is the oxygen self-diffusion coefficient in dry silicate melts, which depends on the viscosity as follows \citep{1996GeCoA..60.4353L}:
\begin{equation}
D = 3.3 \times 10^{-11} {\left( \frac{\eta}{1~\si{Pa}~\si{s}} \right)}^{- 0.722}~\si{m}^{2}~\si{s}^{-1},
\end{equation}
where $\eta$ is the viscosity of melts.
\citet{2022GeCoA.336..104Y} confirmed that this empirical formula is applicable to the oxygen self-diffusion in dry CAI melt.
Then $\tau_{\rm diff}$ is given by
\begin{equation}
\tau_{\rm diff} = 1.4 \times 10^{3} {\left( \frac{r}{0.5~\si{mm}} \right)}^{2} {\left( \frac{\eta}{0.1~\si{Pa}~\si{s}} \right)}^{0.722}~\si{s}.
\end{equation}
The temperature dependence of the viscosity of average chondrule melt is calculated by \citet{2015Icar..254...56H} as follows:
\begin{equation}
\log {\left( \frac{\eta}{1~\si{Pa}~\si{s}} \right)} = - 4.55 + \frac{5084.9~\si{K}}{T - 584.9~\si{K}},
\end{equation}
and $\eta = 0.11~\si{Pa}~\si{s}$ at $T = 2000~\si{K}$.

The relative velocity of ambient gas and silicate melts causes the internal flow within melts \citep[e.g.,][]{2003PThPh.109..717S, 2003EP&S...55..493U}.
The mixing timescale due to the forced convection, $\tau_{\rm mix}$, is given as follows \citep{2003EP&S...55..493U}:
\begin{eqnarray}
\tau_{\rm mix} & \simeq & 5 \times 10^{3} {\left( \frac{\eta}{0.1~\si{Pa}~\si{s}} \right)} {\left( \frac{\rho_{\rm gas}}{10^{-9}~\si{g}~\si{cm}^{-3}} \right)}^{-1} \nonumber \\
               &        & \cdot {\left( \frac{v_{\rm rel}}{10^{2}~\si{m}~\si{s}^{-1}} \right)}^{-2}~\si{s},
\end{eqnarray}
and $\tau_{\rm mix}$ and $\tau_{\rm diff}$ might be of the same order of magnitude if $v_{\rm rel} \sim 10^{2}~\si{m}~\si{s}^{-1}$.
In other words, the interior of molten chondrules would be homogenized by both convective flows and molecular diffusion.

The timescale of homogenization, $\tau_{\rm homo}$, would be approximately given by
\begin{equation}
\tau_{\rm homo} \simeq \min{\left( \tau_{\rm diff}, \tau_{\rm mix} \right)}.
\end{equation}
The dashed lines in Figure \ref{fig6} represent $\tau_{\rm homo}$ as functions of $v_{\rm rel}$.
Here, we assume that the nebular gas is predominantly composed of ${\rm H}_{2}$ gas, and the partial pressure of ${\rm H}_{2}$ gas ($P_{{\rm H}_{2}}$) is $10^{3}$ times higher than that of ${\rm H}_{2}{\rm O}$ vapor \citep[e.g.,][and references therein]{2022GeCoA.336..104Y}.
Then $\rho_{\rm gas}$ is given by
\begin{eqnarray}
\rho_{\rm gas} & = & \frac{m_{{\rm H}_{2}}}{k_{\rm B} T} P_{{\rm H}_{2}} \nonumber \\
               & = & 1.2 \times 10^{-8} {\left( \frac{T}{2000~\si{K}} \right)}^{-1} {\left( \frac{P_{{\rm H}_{2}}}{10^{2}~\si{Pa}} \right)}~\si{g.cm^{-3}},
\end{eqnarray}
where $m_{{\rm H}_{2}} = 2 m_{\rm H}$ is the mass of ${\rm H}_{2}$ molecules.
As $\tau_{\rm homo} \ll \tau$ for $P_{{\rm H}_{2}} \lesssim 10^{2}~\si{Pa}$ (or $P_{\rm water} \lesssim 10^{-1}~\si{Pa}$), we could approximate that the oxygen isotope composition within a chondrule is homogeneous during the isotope exchange reaction.

In contrast, if $P_{\rm water} \gtrsim 1~\si{Pa}$, the timescale of homogenization might be comparable with (or longer than) that of oxygen isotope exchange with ambient water vapor at the surface of melts.
When $\tau_{\rm homo} \gg \tau$, the timescale of isotope exchange between silicate melts and ambient water vapor is limited by the homogenization process instead of the influx of ${\rm H}_{2}{\rm O}$ molecules at the surface of melts \citep[e.g.,][]{2021GeCoA.314..108Y}.

\subsection{Impacts of carbon monoxide on isotope exchange}

Oxygen atoms are included as various kinds of molecules such as ${\rm H}_{2}{\rm O}$, ${\rm C}{\rm O}$, ${\rm C}{\rm O}_{2}$, ${\rm C}{\rm H}_{3}{\rm O}{\rm H}$, silicates, organics, and so on \citep[e.g.,][]{2022arXiv220310863N}.
It is usually thought that ${\rm H}_{2}{\rm O}$, ${\rm C}{\rm O}$, and silicates are major reservoirs of oxygen atoms, and the molar abundance of ${\rm C}{\rm O}$ in the solar nebula would be comparable with that of ${\rm H}_{2}{\rm O}$ \citep[e.g.,][]{1993GeCoA..57.2377W, 2021PhR...893....1O}, although their abundances significantly vary with time and space \citep[e.g.,][]{2020ApJ...899..134K}.
Thus the impacts of carbon monoxide on the evolution of oxygen isotope compositions of chondrules should be discussed.

Laboratory experiments for the ${\rm C}{\rm O}$ gas--silicate melt isotopic exchange by \citet{2022GeCoA.336..104Y} revealed that the isotope exchange efficiency of colliding carbon monoxide molecules is low by three orders of magnitude ($\beta \simeq 3.3 \times 10^{-4}$; see Equation (\ref{eq:beta})), and it is expected that the timescale of oxygen isotope exchange between molten chondrules and ${\rm C}{\rm O}$ gas is also orders of magnitude longer than that for ${\rm H}_{2}{\rm O}$.
In contrast, oxygen isotope exchange between ${\rm C}{\rm O}$ and ${\rm H}_{2}{\rm O}$ gases is orders of magnitude faster compared with isotope exchange between melt and ${\rm H}_{2}{\rm O}$ vapor \citep[see Equation (5) of][]{2004GeCoA..68.3943A}.
Therefore, ${\rm C}{\rm O}$ and ${\rm H}_{2}{\rm O}$ in vapor should have the same oxygen isotope composition during the oxygen isotope exchange between molten chondrules and ambient vapor, as in the case of the heating during igneous CAI formation \citep{2021GeCoA.314..108Y, 2022GeCoA.336..104Y}.
This means the timescale of oxygen isotope exchange would be controlled by the rate of isotope exchange between melt and ${\rm H}_{2}{\rm O}$ vapor.

We note that ${\rm C}{\rm O}$ gas no longer presented in chondrule formation environment if chondrules were formed under highly oxidized condition.
In such case, we might need to discuss the contribution of ${\rm C}{\rm O}_{2}$ gas.

\subsection{Evaporation and recondensation of silicate}
\label{sec:evaporation}

Mass-dependent isotopic fractionation could be caused by evaporation of silicate.
However, the bulk chondrule ${\rm Si}$, ${\rm Mg}$, and ${\rm Fe}$ isotope compositions exhibit small mass-dependent variations \citep[e.g.,][]{2005ASPC..341..432D, 2018crpd.book...91H}, and the relative deviations from the standard is typically $\pm 1 \tcperthousand$.
The impacts of evaporation/condensation on the evolution of bulk oxygen isotopic composition of chondrules could be limited similar to those for the bulk ${\rm Si}$, ${\rm Mg}$, and ${\rm Fe}$ because they evaporated/condensed simultaneously.\footnote{
We note that in situ isotopic analyses for ${\rm Si}$ \citep[e.g.,][]{2020E&PSL.54216318V} and ${\rm Mg}$ \citep[e.g.,][]{2013GeCoA.109..280U} have reported the presence of $\tcperthousand$-level fractionations in Type I chondrule olivine grains.}${}^{,}$\footnote{
We also note that the kinetics of evaporation and condensation of silicates under various ambient gases is still poorly understood in laboratories \citep[e.g.,][]{2015ApJS..218....2T}.
Future experimental studies are necessary for quantitative discussions.}
There are two explanations to prevent evaporation of heated chondrules; one is the formation of chondrules in dust-rich environments \citep[e.g.,][]{2008Sci...320.1617A}, and the other is that chondrule formation via very rapid heating and cooling \citep[e.g.,][]{2012M&PSA..75.5083W}.

On the theoretical side, several studies modeled isotopic fractionation during evaporation and following recondensation.
\citet{2004GeCoA..68.3943A} developed a kinetic evaporation--condensation model to investigate chemical and isotopic fractionation of major elements including ${\rm Mg}$, ${\rm Si}$, and ${\rm O}$.
\citet{2012M&PS...47.1209N} also numerically investigated the role of oxygen isotope exchange reaction during evaporation and recondensation of silicate melt.
These studies found that evolution of oxygen isotopic fractionation must be decoupled from the chemical fractionation of silicate melt, and the trend of PCM line should be created via oxygen isotope exchange reaction between ${\rm H}_{2}{\rm O}$ and molten chondrules.
\citet{2004GeCoA..68.3943A} noted that the absence of large mass-dependent fractionations in all major elements in chondrules is consistent with the situation that chondrules were formed in equilibrium with their environment.
The deviation from the PCM line becomes small when the isotope exchange efficiency is sufficiently high \citep{2012M&PS...47.1209N}.

We note that the exchange reaction that they considered are not identical to that we considered.
\citet{2004GeCoA..68.3943A} considered the reaction ${{\rm M}{\rm O}}_{( {\rm l} )} + {\rm H}_{2} = {\rm M}_{( {\rm g} )} + {\rm H}_{2}{\rm O}$, where ${\rm M}$ denotes ${\rm Mg}$, ${\rm Fe}$, or ${\rm Si}{\rm O}$.
Direct exchange reaction, ${}^{16}{\rm O}_{( {\rm l} )} + {{\rm H}_{2} {}^{i}{\rm O}}_{( {\rm g} )} = {}^{i}{\rm O}_{( {\rm l} )} + {{\rm H}_{2} {}^{16}{\rm O}}_{( {\rm g} )}$ ($i = 17$ or $18$), was not considered in \citet{2004GeCoA..68.3943A}.
This is just because the isotope exchange efficiency was unknown before the laboratory determination by \citet{2021GeCoA.314..108Y}.
\citet{2012M&PS...47.1209N} considered the alternative reaction of direct exchange between water vapor and silicate melts, ${}^{16}{\rm O}_{( {\rm l} )} + {}^{i}{\rm O}_{( {\rm g} )} = {}^{i}{\rm O}_{( {\rm l} )} + {}^{16}{\rm O}_{( {\rm g} )}$, because hydrogen was not considered in the system.
The isotope exchange efficiency was treated as a free parameter.
As $c_{\rm s}$ of ${\rm O}$ and ${\rm H}_{2}{\rm O}$ molecules is approximately the same, \citet{2012M&PS...47.1209N} mentioned that the exact treatment of gas species of oxygen may not be critical for qualitative arguments.

We also note that hydrogen could play a major role in the evaporation of chondrules \citep[e.g.,][]{1996GeCoA..60.1445N, 2002Icar..160..258M}.
Future studies on the detailed modeling of oxygen isotope exchange during evaporation/recondensation in nebular hydrogen gas would be essential.

\subsection{Oxidization/reduction of iron in chondrules}
\label{sec:oxidization}

Chondrules exhibit wide ranges of Mg\# values (i.e., the molar percentage of MgO / (MgO + FeO)), and they are divided into two groups: Type I (Mg\# $> 90$) and Type II (Mg\# $< 90$).
It is known that the abundance of iron metals in Type II chondrules is significantly lower than that in Type I chondrules in both ordinary and carbonaceous chondrites \citep[e.g.,][]{2015GeCoA.160..277V}.

The origin of iron metals in chondrules has been debated so far.
A plausible scenario is that metals were produced by melting and coalescence of preexisting fine metals during chondrule formation \citep[e.g.,][]{2013M&PS...48.1981J}.
Other scenarios such as capture of colliding metals \citep[e.g.,][]{2022GeCoA.319..254N} and reduction of FeO \citep[e.g.,][]{2001GeCoA..65.4567C, 2003M&PS...38...81L} are also proposed as the origin of metal grains within chondrules.
\citet{2010PhDT.......453B} performed detailed electron microprobe studies to determine the bulk composition of chondrules in both ordinary and carbonaceous chondrites, and there was no significant difference in the bulk Fe content between Type I and II chondrules.
This finding might indicate that oxidization/reduction of Fe plays a major role in the origin of dichotomy for the abundance of iron metals in Type I and II chondrules.

The bulk Fe content of Type II chondrules in CR and CO chondrites are approximately 20\% in mass \citep[e.g.,][]{2010PhDT.......453B}.
If we assume that all Fe in Type II chondrules exists as FeO and they were formed via oxidation of Fe metals, the number of oxygen atoms in chondrules might increase by 10--20\% during oxidation, depending on the initial chemical composition.
The possible impacts of oxidization/reduction of Fe on the evolution of oxygen isotope composition should be investigated in future studies.\footnote{
When oxidization/reduction reactions finish before the completion of gas--melt isotope exchange reactions, we expect that oxidization/reduction reactions barely affect the final oxygen isotopic compositions.
}

\subsection{Chondrules in ordinary chondrites}

In this paper, we applied our theoretical model to constraining the formation mechanisms of chondrules in carbonaceous chondrites.
In this section, we briefly discuss the formation process of chondrules in ordinary chondrites.

Chondrules in carbonaceous chondrites are thought to be formed in dust-rich environments \citep[e.g.,][]{2010GeCoA..74.4807R}.
For ordinary chondrites, \citet{2008Sci...320.1617A} found that the abundance of sodium remained approximately constant during chondrule formation, and evaporation would be prevented when chondrules were molten.
Thus chondrules in ordinary chondrites would also be formed in dust-rich environments.

\citet{2021GeCoA.313..295P} performed in situ oxygen isotopic analyses of Type I olivine-rich chondrules in LL3 ordinary chondrites.
They found the following features: (1) $^{16}{\rm O}$-rich relict olivine grains were identified in chondrules, although they are much less abundant than in chondrules in carbonaceous chondrites, (2) oxygen isotopic compositions of relict grains vary widely along the PCM line,\footnote{
\citet{2021GeCoA.313..295P} also reported a small but systematic offset from the PCM line.
} and (3) host olivine grains exhibit small mass-dependent isotopic variations within individual chondrules ($\sim 1\tcperthousand$ for $\delta{}^{18}{\rm O}$, 1 standard deviation).
\citet{2021GeCoA.313..295P} noted that host olivine grains in carbonaceous chondrite chondrules also exhibit small intra-chondrule mass-dependent variations, with $\sim 1\tcperthousand$ for $\delta{}^{18}{\rm O}$.
They concluded that the similar features observed in ordinary and carbonaceous chondrite chondrules would be the evidence that the mechanism for chondrule formation could be the same.
Our results indicate that the small mass-dependent isotopic variations within individual chondrules in ordinary chondrules would be inconsistent with chondrule formation via shock-wave heating around planetesimals (Section \ref{sec:shock}).

\subsection{Oxygen isotope ratio of igneous calcium--aluminum-rich inclusions}

We note that our theoretical model for isotope exchange with spheres and ambient vapor is applicable for many problems in geo- and cosmochemistry.
The oxygen isotope ratio of igneous CAIs in carbonaceous chondrites is also plotted on the mixing line called the carbonaceous chondrite anhydrous mineral line \citep{1977E&PSL..34..209C}, and their oxygen isotopic composition is thought to evolve via exchange reaction with ambient vapor \citep[e.g.,][]{1998Sci...282.1874Y, 2016E&PSL.440...62A, 2018GeCoA.221..318K}.\footnote{
Oxygen isotope exchange during secondary alteration on the parent body could have also affect the oxygen isotope ratio of CAIs \citep[e.g.,][]{2019GeCoA.246..419K}.
However, experimental studies \citep[e.g.,][]{1994GeCoA..58.3713R} reported that the oxygen diffusivity in melilite grains is sufficiently low, and the effect of secondary alteration on the parent body would be limited in $\si{\micro m}$-sized grain scale \citep[see also][]{2021GeCoA.314..108Y}.}
Our results suggest that not only chondrules but also igneous CAIs would be formed via mechanisms that do not trigger large relative velocity between silicate melts and ambient gas.
This would be an important constraint on the unknown heating mechanism of igneous CAIs.
In addition, the cooling rate of igneous CAIs was evaluated from the spatial heterogeneity of oxygen isotopic composition within a CAI \citep[e.g.,][]{2021M&PS...56.1224K, 2021GeCoA.314..108Y}.
\citet{2021GeCoA.314..108Y} concluded that the duration of heating of typical igneous CAIs would be a few days or longer, which is comparable to the Kepler timescale at the location of CAI formation ($\sim 10~{\rm days}$ at $10^{-1}~{\rm au}$ from the Sun).\footnote{
It should be noted that CAIs did not necessarily form at $\sim 10^{-1}~{\rm au}$ from the Sun; they might have condensed at $\sim 1~{\rm au}$ \citep[e.g.,][]{2019A&A...624A.131J}.}
Thus we can speculate that the formation of igneous CAIs could be linked with the star--disk interaction \citep[e.g., protostellar flares and disk winds;][]{2019ApJ...878L..10T, 2022ApJ...941...73T}.
Dust dynamics and thermal history in the innermost region of the solar nebula should be investigated from these points of view in future studies.

\section{Summary}

Oxygen isotope compositions of chondrules have evolved via isotope exchange reaction with ambient vapor, and those would reflect the chondrule-forming environment such as the dust-to-gas and water-to-rock ratios of the environment.
Oxygen isotope exchange reaction between chondrule precursors and ambient vapor is regarded as the origin of the variation of oxygen isotope compositions of chondrules.
The isotope exchange efficiency between silicate melt and water vapor was determined by \citet{2021GeCoA.314..108Y}, and we can evaluate the timescale of isotope exchange reaction.
However, there was no theoretical model of isotope exchange reaction that is applicable for moving chondrule melt in ambient vapor.
The relative velocity between melt and ambient vapor must affect the isotope exchange reaction; this is because the influx of colliding molecules at the chondrule surface depends on the relative velocity.
As shock-wave heating around planetesimals/protoplanets with high heliocentric eccentricity is one of the best-studied mechanisms for chondrule formation, we need to consider the effect of relative velocity on the oxygen isotope exchange reaction.

In this study, we develop a theoretical model of isotope exchange reaction between spheres and ambient vapor with nonzero relative velocities.
We derived a modified equation of the flux of ambient vapor that is applicable for moving spheres (Section \ref{sec:influx}), and we revisited the relation between the influx and efflux in isotope exchange reactions based on the concept of the isotope fractionation at the equilibrium (Section \ref{sec:efflux}).

We found that large mass-dependent fractionation would be caused by isotope exchange with ambient vapor unless the relative velocity is sufficiently smaller than $c_{\rm s}$ of reacting molecules (Section \ref{sec:results}).
Our findings indicate that the relative velocity between chondrules and ambient water vapor would be lower than several $100~\si{m}~\si{s}^{-1}$ when chondrules crystallized.


We acknowledge that the calculations have been performed assuming no net transport of oxygen atoms, so the kinetic effects of evaporation/recondensation are yet to be worked out.
We will develop our theoretical model further in future studies.

\section*{Acknowledgments}

We thank two anonymous reviewers for thoughtful comments.
S.A. was supported by the research grant for JAMSTEC Young Research Fellow.
This work was supported by JSPS KAKENHI grant numbers JP20H04621 and JP22K18741 (T.U.).

\clearpage
\appendix

\section{Derivation of Equation (\ref{eq:vrel})}
\label{app:vrel}

For simplicity, we consider a one-dimensional normal shock model as in earlier studies \citep[e.g.,][]{2001Icar..153..430I, 2002M&PS...37..183D, 2005Icar..175..289M, 2014ApJ...797...30J, 2023ApJ...948...73M}.
We do not perform hydrodynamic simulations but assume a simple gas structure as in previous studies \citep{2019ApJ...877...84A, 2022ApJ...927..188A}.
The gas velocity with respect to the shock front, $v_{\rm gas}$, evolves as a function of the distance from the shock front, $x$.
We set that the pre-shock gas velocity is $v_{0}$ and the post-shock gas velocity is $v_{\rm post}$.
We can imagine that $v_{\rm gas} \to v_{0}$ for $x \to \infty$ when shocks are caused by planetesimals.
Here we assume that $v_{\rm gas}$ is given as follows:
\begin{equation}
v_{\rm gas} = 
\begin{cases}
\displaystyle v_{0} & {( x < 0 )}, \\
\displaystyle v_{0} + {\left( v_{\rm post} - v_{0} \right)} \exp{\left( {- x}/{L} \right)} & {( x \geq 0 )},
\end{cases}
\label{eq:vg}
\end{equation}
where $L$ denotes the spatial scale of the shock.
In the post-shock region of $x > 0$, the rate of the gas velocity change per unit length, ${{\rm d}v_{\rm gas}} / {{\rm d}x}$, is given by
\begin{equation}
{\left| \frac{{\rm d}v_{\rm gas}}{{\rm d}x} \right|} = \frac{\left| v_{\rm gas} - v_{0} \right|}{L}.
\label{eq:dvgdx}
\end{equation}

The velocity of chondrules with respect to the shock front, $v$, evolves due to the gas drag \citep[e.g.,][]{1991Icar...93..259H}.
We assume that the temperature of chondrules is equal to that of the gas for simplicity.
The rate of the chondrule velocity change per unit length, ${{\rm d}v} / {{\rm d}x}$, is given as follows \citep[e.g.,][]{2019ApJ...877...84A, 2022ApJ...927..188A}:
\begin{equation}
\frac{4 \pi}{3} r^{3} \rho \frac{{\rm d}v}{{\rm d}x} = - \frac{C_{\rm D}}{2} \pi r^{2} \rho_{\rm gas} \frac{\left| v - v_{\rm g} \right|}{v} {\left( v - v_{\rm gas} \right)},
\label{eq:v}
\end{equation}
where $C_{\rm D}$ is the drag coefficient.
For the subsonic limit, $C_{\rm D}$ is given by the following equation \citep[e.g.,][]{2014ApJ...797...30J}:
\begin{equation}
C_{\rm D} = \frac{16}{3 \sqrt{\pi}} {\left( 1 + \frac{\pi}{8} \right)} {s_{\rm gas}}^{-1},
\end{equation}
where
\begin{equation}
s_{\rm gas} = \frac{{\left| v - v_{\rm gas} \right|}}{c_{\rm gas}}.
\end{equation}
Then we can rewrite Equation (\ref{eq:v}) as follows:
\begin{equation}
{\left| \frac{{\rm d}v}{{\rm d}x} \right|} \simeq \frac{1}{0.64} \frac{\rho_{\rm gas}}{\rho} \frac{c_{\rm gas}}{v} \frac{{\left| v - v_{\rm gas} \right|}}{r}.
\label{eq:dvdx}
\end{equation}

The relative velocity between a chondrule and gas, $v_{\rm rel}$, is defined as 
\begin{equation}
v_{\rm rel} \equiv {\left| v - v_{\rm gas} \right|}.
\end{equation}
When $L$ is larger than the chondrules' stopping length, $l_{\rm stop}$, chondrules would be dynamically coupled with gas at $x \gg l_{\rm stop}$.
Then, the following relations,
\begin{eqnarray}
v                                          & \simeq & v_{\rm gas}, \\
{\left| \frac{{\rm d}v}{{\rm d}x} \right|} & \simeq & {\left| \frac{{\rm d}v_{\rm gas}}{{\rm d}x} \right|},
\end{eqnarray}
could be satisfied at $x \gg l_{\rm stop}$, and the dynamics of chondrules is in quasi-steady state.
We note that the dynamical coupling is not complete when we consider finite values of $l_{\rm stop}$ and $L$.
In this case, we can evaluate $v_{\rm rel}$ by combining Equations (\ref{eq:dvgdx}) and (\ref{eq:dvdx}) as follows:
\begin{equation}
v_{\rm rel} \simeq 0.64 \frac{\rho}{\rho_{\rm gas}} \frac{v_{\rm gas}}{c_{\rm gas}} \frac{\left| v_{\rm gas} - v_{0} \right|}{L} r.
\label{eq:steady}
\end{equation}
It is clear that $v_{\rm rel}$ is inversely proportional to $L$ and $v_{\rm rel} \to 0$ for $L \to \infty$.
In addition, $v_{\rm rel} \to 0$ for $x \to \infty$ because ${\left| v_{\rm gas} - v_{0} \right|} = 0$ at $x = \infty$.

We can derive Equation (\ref{eq:steady}) by another way.
As the velocity of chondrules evolves due to the gas drag, the equation of motion of chondrules is given by
\begin{equation}
{\left| a \right|} = \frac{v_{\rm rel}}{t_{\rm stop}},
\end{equation}
where $a$ is the acceleration of chondrules and 
\begin{equation}
t_{\rm stop} = 0.64 \frac{\rho}{\rho_{\rm gas}} \frac{r}{c_{\rm gas}},
\end{equation}
is the stopping time of chondrules.
In steady-state, we can assume $a \simeq a_{\rm gas}$, where $a_{\rm gas}$ is the acceleration of the gas which is given by
\begin{equation}
{\left| a_{\rm gas} \right|} = \frac{\left| v_{\rm gas} - v_{0} \right|}{L} v_{\rm gas}.
\end{equation}
Then we can obtain the following equation:
\begin{equation}
v_{\rm rel} \simeq {\left| a_{\rm gas} \right|} t_{\rm stop},
\end{equation}
and this is identical to Equation (\ref{eq:steady}).
Assuming that the nebular gas is ${\rm H}_{2}$ gas whose temperature is $2000~{\rm K}$, $v_{0} = 10~{\rm km}~{\rm s}^{-1}$, $v_{\rm gas} = 0.2 v_{0}$, and $\rho = 3~{\rm g}~{\rm cm}^{-3}$, we obtain the following values of $a_{\rm gas}$ and $t_{\rm stop}$:
\begin{equation}
a_{\rm gas} = 1.6 \times 10^{2}~{\left( \frac{L}{10^{4}~{\rm km}} \right)}^{-1}~\si{cm.s^{-2}},
\end{equation}
and
\begin{equation}
t_{\rm stop} = 2.4 \times 10^{2}~{\left( \frac{\rho_{\rm gas}}{10^{-9}~{\rm g}~{\rm cm}^{-3}} \right)}^{-1} {\biggl( \frac{r}{0.5~{\rm mm}} \biggr)}~\si{s}.
\end{equation}

We note that the evaluation of $v_{\rm rel}$ in Equation (\ref{eq:steady}) is not more than an order-of-magnitude estimate.
The spatial variation of $v_{\rm gas}$ should be more complicated than that assumed in Equation (\ref{eq:vg}), and bow shocks around planetesimals have three-dimensional structure in reality.
Thus quantitative discussions are left for future work.

\clearpage

 \bibliographystyle{elsarticle-harv} 
 \bibliography{sample631}





\end{document}